\documentclass[doublecol]{epl2}

\newcommand{\be}{\begin{equation}}
\newcommand{\ee}{\end{equation}}
\newcommand{\bs}{\begin{subequations}}
\newcommand{\es}{\end{subequations}}

\title{Lithium atom storage in nanoporous cellulose via surface induced ${\rm{L}}{{\rm{i}}_2}$ breakage}
\shorttitle{Surface induced ${\rm{L}}{{\rm{i}}_2}$ breakage}

\author{Mathias Bostr{\"o}m\inst{1,2}\and Dan Huang \inst{1} \and Weihong Yang \inst{1} \and Clas Persson\inst{1,2,3}  \and Bo E. Sernelius\inst{4}}

\institute{ \inst{1}Dept of Materials Science and Engineering, Royal Institute of Technology, SE-100 44 Stockholm, Sweden, EU\\
  \inst{2}Centre for Materials Science and Nanotechnology, University of Oslo, P. O. Box 1048 Blindern, NO-0316 Oslo, Norway\\
 \inst{3}Department of Physics, University of Oslo, P. O. Box 1048 Blindern, NO-0316 Oslo, Norway\\
 \inst{4}Division of Theory and Modeling, Department of Physics,  Chemistry and Biology, Link\"{o}ping University, SE-581 83 Link\"{o}ping, Sweden, EU}
 \shortauthor{M. Bostr{\"o}m \etal}

\abstract{
We demonstrate a physical mechanism that enhances a splitting of diatomic  ${\rm{L}}{{\rm{i}}_2}$ at cellulose surfaces.  
The origin of this splitting is a possible surface induced diatomic excited state resonance repulsion. 
The atomic Li is then free to form either physical or chemical bonds with the cellulose surface and 
even diffuse into the cellulose layer structure.  This allows for an enhanced storage capacity of atomic Li in nanoporous cellulose. }
\begin{document}
\maketitle


Recently, there has been considerable interest, both theoretically and experimentally, in fluctuation induced forces.
The zero-point and thermal fluctuations of the electromagnetic field in the vicinity of a molecule, or at surfaces, are different from those in free space. This affects the intermolecular forces and the changes can be attractive or repulsive.\,\cite{Sush,Zwol1,Feiler}  
Notably, the influence of a surface can be to turn the first order dispersion potential (also known as resonance potential) between two atoms from attraction to repulsion. Resonance interaction occurs between two atoms in an excited state.

Biologically based electronics is an emerging field that aims to
 design and manufacture inexpensive, easy to recycle and flexible
 devices with a complex functionality. Paper with its long thin
 cellulose fibers has the advantage of a large effective surface area
 that can thereby potentially be used to store energy. Cellulose
 exhibits excellent characteristics, like hydrophilicity,
 biodegradability, biocompatibility, and capacity for chemical
 modifications to change the surface properties. Thereby cellulose has
 potential for chemical storage. Inexpensive and efficient Li-ion
 batteries are considered to be a future power source for portable
 electronic applications and electric vehicles. Paper based storage
 of Li may therefore play a role in future small, hazard free, and
 biodegradable power devices.\,\cite{Hu,Nyst,ChunJMC2012}   The possibility to modify the
 cellulose surface properties is an advantage for tailoring specific
 properties. For instance, it has been suggested that nanostructures
 near cellulose surfaces can lead to enhanced Li storage.\,\cite{Guo,Yue}

\begin{figure}
\includegraphics[width=7cm]{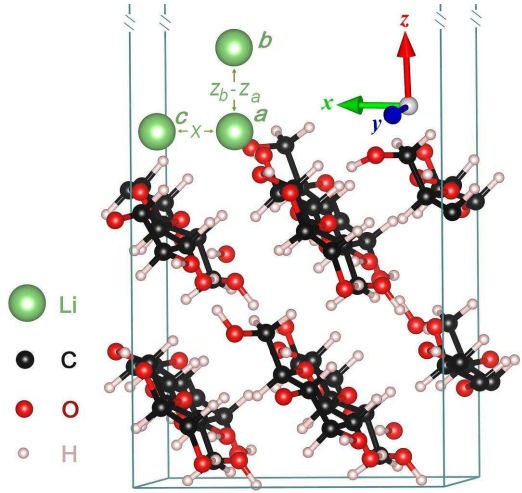}
\caption{(Color online) Two examples of  diatomic ${\rm{L}}{{\rm{i}}_2}$  adsorbed at a (110) cellulose surface. The first example is with 
atoms $a$ and $c$ both at a distance ${z_a = z_c}$ from the surface. The distance between the two atoms is $\rho = x$\,\AA. 
Atomistic DFT calculations have been used to deduce the non-resonant bond enthalpy for this geometry to be 0.12 eV with a 
bond length of 2.69\,\AA. The second example is with Li atoms $a$ and $b$
situated one outside the other near a cellulose surface. The closest atom is at ${z_a}$ from the surface. 
The distance between the two atoms is $\rho = z_b-z_a$\,\AA. 
The non-resonant bond enthalpy from the DFT calculation for this geometry is 0.11 eV with a bond length of 2.81\,\AA.}
\label{figu1}
\end{figure}

In the present letter we focus on the effect of interatomic forces causing breakage of diatomic Li molecules at porous cellulose surface. By employing an atomistic modeling within the density functional theory (DFT) in conjunction with an employed semiclassical theory for the resonance interaction, we demonstrate that resonance interaction\,\cite{Sherkunov,Bostrom1,BostPRA2012}  
between excited-state ${\rm{L}}{{\rm{i}}_2}$ atom pairs near surfaces  can  give rise to sufficient repulsion to cause diatomic 
${\rm{L}}{{\rm{i}}_2}$ breakage. In subsequent events,  when the atomic Li is near a cellulose surface, it can interact with 
the cellulose structure via either physisorption (Casimir-Polder forces) or chemisorption.\,\cite{Xie} It may even diffuse into 
the cellulose layer structure; this leads potentially to new ways for storage of atomic Li in nanoporous cellulose. 
We end with a few conclusions.  Atomic scale illustration of the system considered is shown in  Fig.\,\ref{figu1}.

\begin{figure}
\includegraphics[width=8cm]{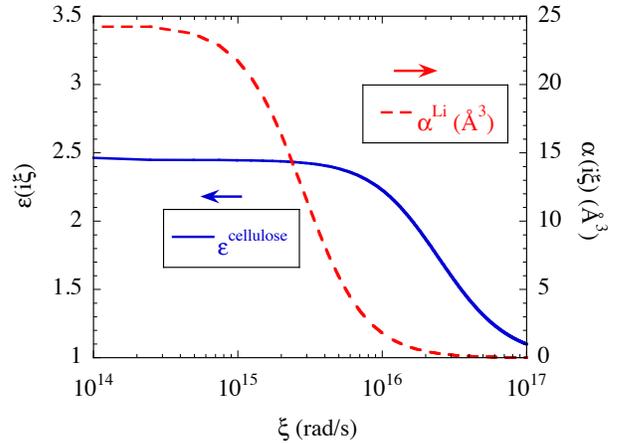}
\caption{(Color online) The dielectric function of cellulose and the polarizability of lithium for imaginary frequencies.}
\label{figu2}
\end{figure}

 Our atomistic modeling of Li on a (110) cellulose surface is based on density functional theory with dispersion correction 
(DFT-D2),\,\cite{Grimme} The surface is constructed by a $\sqrt{2}\,\times \sqrt{2}\,\times 1$ supercell slab with a {35 \AA} vacuum region.
An energy cutoff of 500 eV and a special {\bf k}-point sampling over an $4\,\times 4\,\times 1$ Monkhorst-Pack mesh are used for the surface system. 
The crystal structure of cellulose $I_\beta$ was chosen to be the monoclinic P$112_1$ structure. Our calculated structure parameters for bulk $I_\beta$ ($a$ = 7.64 \AA, $b$ = 8.13 \AA, $c$ = 10.38 \AA, $\alpha$ = $\beta$ = 90$^\circ $, and $\gamma$ = 96.69$^\circ$)
are consistent with the experimental data\,\cite{Nishy} and other calculated results.\,\cite{Buck,Li} 
The imaginary part of the dielectric function is calculated by an independent-particle approximation, in which the wave function is generated by the DFT-D2 method. Thereafter, the spectra on the imaginary frequency axis are obtained via the Kramers-Kronig relation.
The corrected band gap value is estimated to $E_g$ = 8.58 eV by means of the partial self-consistent  $GW_0$ method.\,\cite{Shish,Gonz2010} 
The static ion-clamped dielectric matrix is calculated by using the density functional perturbation theory.\,\cite{Gajd} 
The dielectric function of cellulose and the polarizability of lithium as functions of frequencies,  needed to evaluate the dispersion interactions, are shown in Fig.\,\ref{figu2}. 
For interactions at the short distances considered here only frequencies in the range $10^{14}$ rad/s and upwards give noticeable contributions to the interaction. For cellulose there are additional excitation processes for very low frequencies that leads to an enhancement of the dielectric function for very low $\xi$.  In the figure we have focused on the relevant frequency range for both the dielectric function and for the polarizability.

By means of the DFT, we analyze the non-resonant bond enthalpy and dissociation energy in the following way. 
First, the bond enthalpy of a diatomic ${\rm{L}}{{\rm{i}}_2}$ molecule in free space is obtained from the difference between the total 
energy of ${\rm{L}}{{\rm{i}}_2}$ and that of the two Li atoms separated. It is found to be 0.89 eV for an equilibrium bond length of 2.68\,\AA,
which fairly well agree with experimental values of 1.09 eV and 2.67\,\AA, respectively.\,\cite{Expt1,Expt2} 
Second, the energy for adsorption of sole Li atoms on the surface is determined by comparing the total energy of a Li on the 
surface with the corresponding energies for a free surface plus half the energy of the diatomic Li molecule in free space. 
This single atom is bonded to two oxygen atoms with the bond lengths 1.98\,{\AA}  and 1.99\,{\AA} ({\it i.e.}, atom position $a$ in Fig. 1), and the adsorption energy is found to be $-0.59$ eV. 
Third, the energy for adsorption of a ${\rm{L}}{{\rm{i}}_2}$ molecule is determined by comparing the total energy of a chemically bonded Li pair 
on the surface with the energy of a diatomic ${\rm{L}}{{\rm{i}}_2}$ molecule in free space. 
If the Li pair is oriented along the $z$-direction perpendicular to the surface ({\it i.e.}, atoms $a$ and $b$), 
the adsorption energy is found to be $-1.32$ eV with a Li-Li bond length of 2.81\,\AA. Now, the Li-O bonds are slightly weakened to lengths of 1.98\,{\AA}  and 2.02\,\AA.
However, if the Li pair is oriented along the $x$-direction parallel to the surface ({\it i.e.}, atoms $a$ and $c$), 
the adsorption energy is $-1.31$ eV and the Li-Li bond length is 2.69\,\AA. 
Here, the second Li atom is also bonded to a cellulose oxygen atom with a bond length of 2.06\,\AA.
Forth, we calculate the bond dissociation energy on the surface by comparing the total energy of a bonded Li pair on the surface 
with the energy of two sole Li atoms that both are on the most stable position on the surface ({\it i.e.}, two $a$ atoms). 
If the Li pair is initially along the $z$-direction the dissociation energy is 0.12 eV, and if the Li pair is initially along the 
$x$-direction the energy is 0.11 eV. This positive value indicates that the chemical adsorption of ${\rm{L}}{{\rm{i}}_2}$ on the (110) surface is associative. 
This means that there is a possibility for exothermic ${\rm{L}}{{\rm{i}}_2}$ adhesion to the surface, and that the adsorbed Li pair can be positioned with different orientations. Dissociative diffusion requires an additional repulsive interaction energy of around 0.1 eV.

We will compare the results from the DFT evaluated ground-state 
interactions and the result from semiclassical theory. The Lamb shift as calculated from semiclassical theory has been shown to be identical to the corresponding result from quantum electrodynamics,\,\cite{Mah1974} and the semiclassical theory has 
in many similar cases been shown to give results identical to those of perturbative quantum mechanics.\,\cite{Mah,Pars,Ninhb}  Semi-classical theory furthermore revealed that the underlying theory of resonance interactions in free space derived from perturbative quantum electrodynamics (QED) may be incorrect at large distances.\,\cite{Bostrom1}  Here we consider short distances where different theories give identical results.
We explore if the interaction energy in the case of two identical atoms in an excited state is modified near surfaces.\,\cite{BostPRA2012} 
The resonance conditions in the semiclassical theory 
is here separated into one anti-symmetric and one symmetric part. 
Since the excited symmetric state has a much shorter life time 
than the anti-symmetric state the system can be trapped in an excited anti-symmetric state.  We use a simple  
semiclassical expression for the resonance interaction energy between two identical polarizable atoms excited in the 
$j$ = {\it x-, y-,} or {\it z-}direction (where the {\it z}-direction is perpendicular to the cellulose surface).\,\cite{Bostrom1,BostPRA2012} 
The results for excited state atom-atom interaction are obtained by solving numerically the equations of motion for the excited system. 
The resonance condition\,\cite{Bostrom1} is then determined from the following equation: 
$\tilde 1 -\,\alpha (\omega )^2 {\tilde T^2}(\rho |\omega ) = 0$, where $\tilde T$ is the susceptibility tensor for two atoms near 
a surface\,\cite{Buhmann1,Buhmann2} and   $\alpha(\omega)$  the  polarizability  of a Li atom. The resonance interaction energy 
of this antisymmetric state can be evaluated  from
\begin{equation}
U_j(\rho ) = {\hbar \over {\pi}} \int\limits_{0}^\infty  d\xi \ln [1 + \alpha(i\xi ){T_{jj}}(\rho |i\xi )],
\label{Eq1}
\end{equation}
where $\rho$ is the distance between the two atoms. In the case of anisotropic excitations the leading term in the near zone is $\propto \rho^{-3}$ in good agreement with previous results. However, for isotropic excitations the leading term in the near zone is $\propto \rho^{-6}$, similar to the attractive van der Waals potential between ground-state atoms. 
\begin{figure}
\includegraphics[width=8cm]{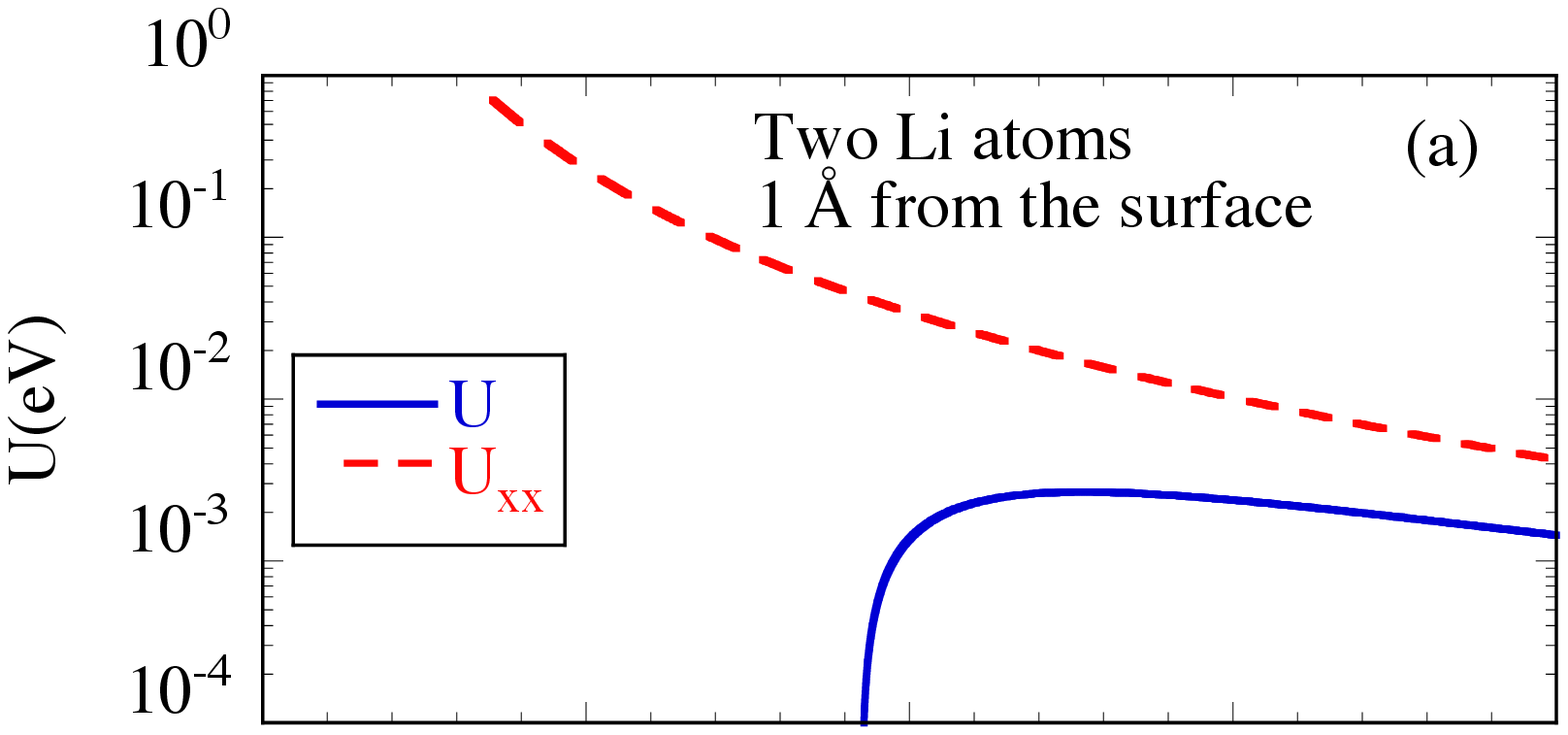}
\includegraphics[width=8cm]{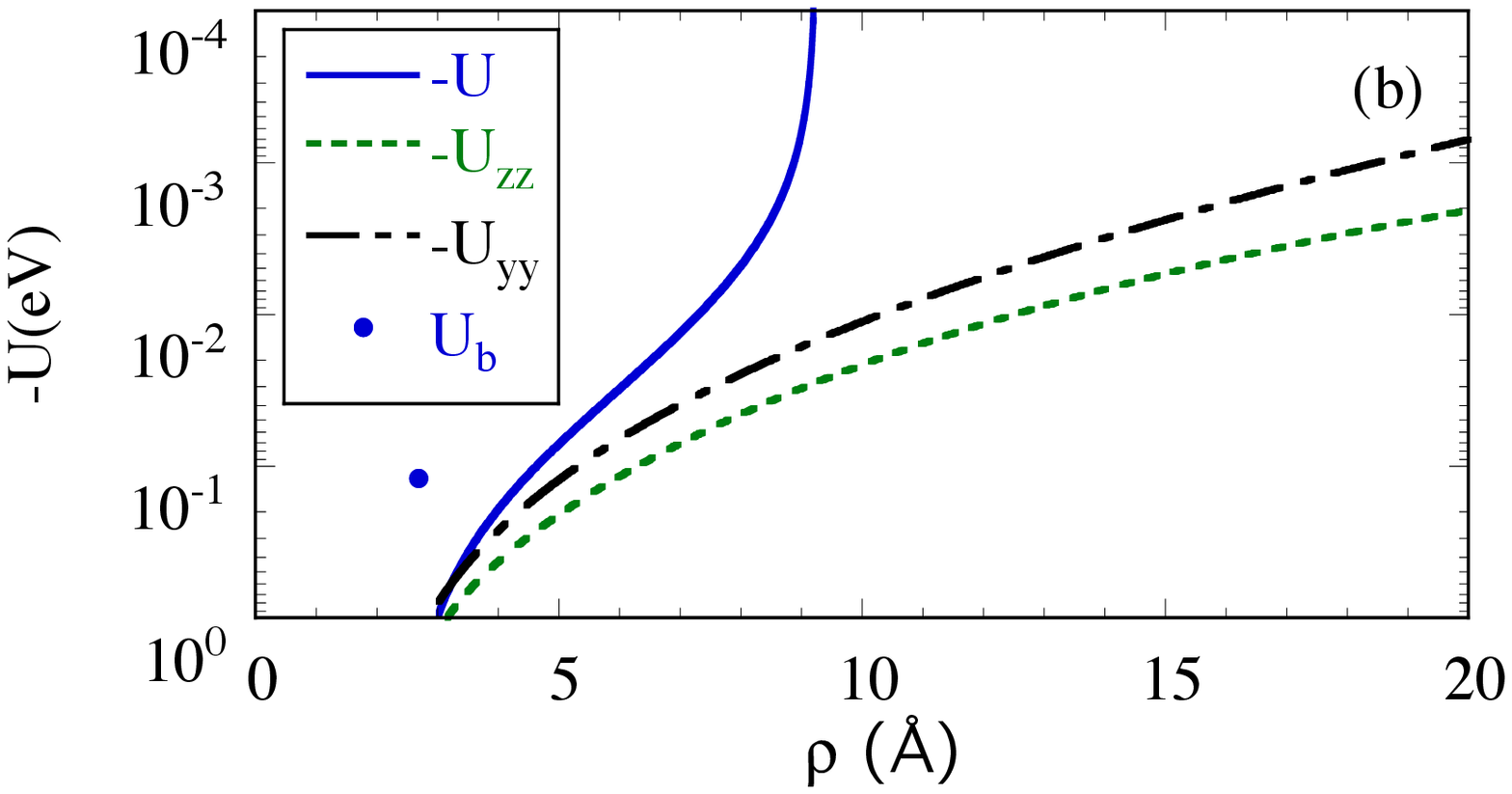}
\caption{(Color online) The resonance interaction between two isotropically excited Li atoms, a distance $\rho$ apart,  
adsorbed on a cellulose surface. The atoms are a distance $z_a = z_c = $ 1 {\AA} from the surface. When the atoms are close to each other an attractive van der Waals interaction ($\propto \rho^{-6}$) dominates the resonance interaction. Atomistic 
DFT calculations give the non-resonant bond enthalpy for this geometry to be 0.12 eV with a bond length of 2.69\,\AA.}
\label{figu3}
\end{figure}

The fact that there can be repulsive and attractive contributions follows already from the basic theory\,\cite{McLone} of simple neutral atoms with dipole moments having a resonance interaction energy in free space between one atom in an excited state and one ground-state atom. An approximation valid in the non-retarded limit is the Coulomb approximation and we denote the electric dipole moment of an atom by ${\bf q}^{(r)}$. If the principal axes are chosen as coordinate axes the dipole-dipole tensor becomes diagonal. If we further choose the $z$-axis to point along the line connecting the two atoms we have
\begin{equation}
U(\rho ) = {U_x} + {U_y} + {U_z} = \frac{{q_x^{(1)}q_x^{(2)}}}{{{\rho ^3}}} + \frac{{q_y^{(1)}q_y^{(2)}}}{{{\rho ^3}}} - 2\frac{{q_z^{(1)}q_z^{(2)}}}{{{\rho ^3}}}.
\label{Eq2}
\end{equation} Depending on the orientation of the system there can be an attractive or repulsive resonance interaction. These results are correct for atoms in free space at not too large separations. The resonance interaction at large separations has  for the different orientations   been written as $U^{\Pi/\Sigma}(\rho)=-f^{\Pi/\Sigma} (\rho) C_3^{\Pi/\Sigma}/\rho^3$ where $C_3^{\Sigma}=-2 C_3^{\Pi}$.\,\cite{WangPRA}  This result was shown by Ninham and co-workers to be questionable in the retarded limit.\,\cite{Bostrom1} We will here use theories that give an accurate treatment of resonance interaction between two atoms that are close to each other, especially when they are near a surface (and in the last figure accounting also for finite atomic size). For completeness we consider both isotropically excited states and system excited parallel or perpendicular relative to a line joining the two atoms. The largest repulsion comes for excitation parallel to a line joining the two atoms and this will be seen to be large enough to enable breaking of  the surface adsorbed atom pairs.
It is notable that retardation is irrelevant at these separations for anisotropic excitations. However, due to large cancellations of {\it x-, y-}, and {\it z-}contributions retardation plays some role for isotropically excited atom pairs.

The binding energy of diatomic ${\rm{L}}{{\rm{i}}_2}$  in free space from atomic DFT calculations is around 0.89 eV with a bond length of 2.69\,\AA. 
For ground-state and isotropically excited atom pairs the interaction in vacuum is attractive for all separations.  
There exists only a small possibility to have  bond splitting of  ${\rm{L}}{{\rm{i}}_2}$  in  free space and that is only for a system excited in the  direction defined by a line joining the two atoms.  We show in Fig.\,\ref{figu3} the interaction between two Li atoms in an isotropically excited state, both adsorbed on a cellulose surface.  The corresponding contributions from excitations in the 
$x$-, $y$-, and $z$-directions are also shown in  Fig.\,\ref{figu3}. Only distances beyond the bond length (of the order 3\,\AA) 
are shown since for shorter separations chemical interactions dominate. For larger separations bond splitting via resonance 
interaction can occur. We used atomistic scale DFT calculations to deduce that the non-resonant bond enthalpy (energy of adsorption) for this geometry is  0.12 eV.  We see that repulsive contributions can be substantial only if the atoms have an excitation in the direction defined 
by a line joining the two atoms. For such excitation the resonance repulsion (around 1 eV) is sufficient to break the bond between the atom pair. 
The effect of physical adsorption, of one or both of the atoms, on the surface is to change the interaction between excited-state Li atom pairs. This is accounted for in our calculations via the changes in the Green's functions that describe the coupled system (surface, atoms and the electromagnetic field).

\begin{figure}
\includegraphics[width=8cm]{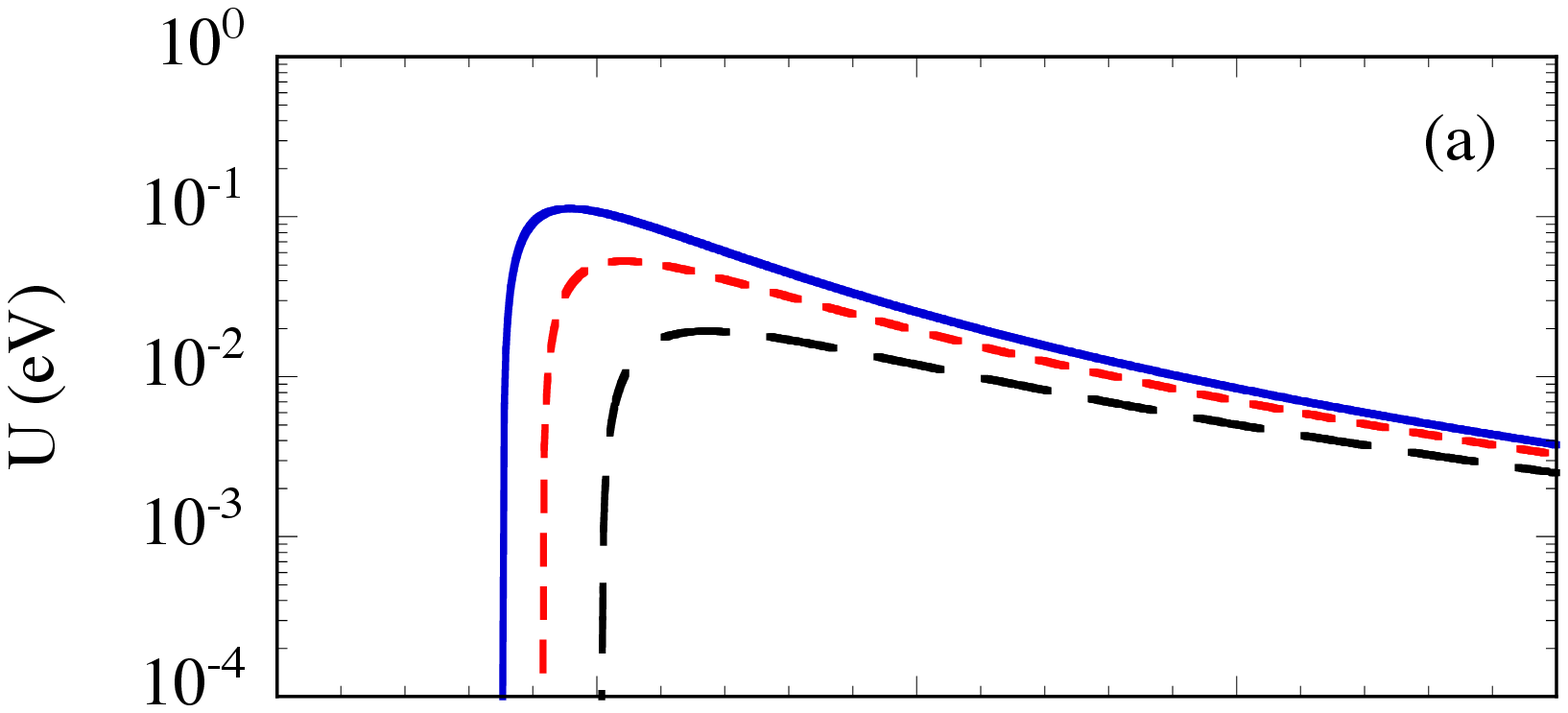}
\includegraphics[width=8cm]{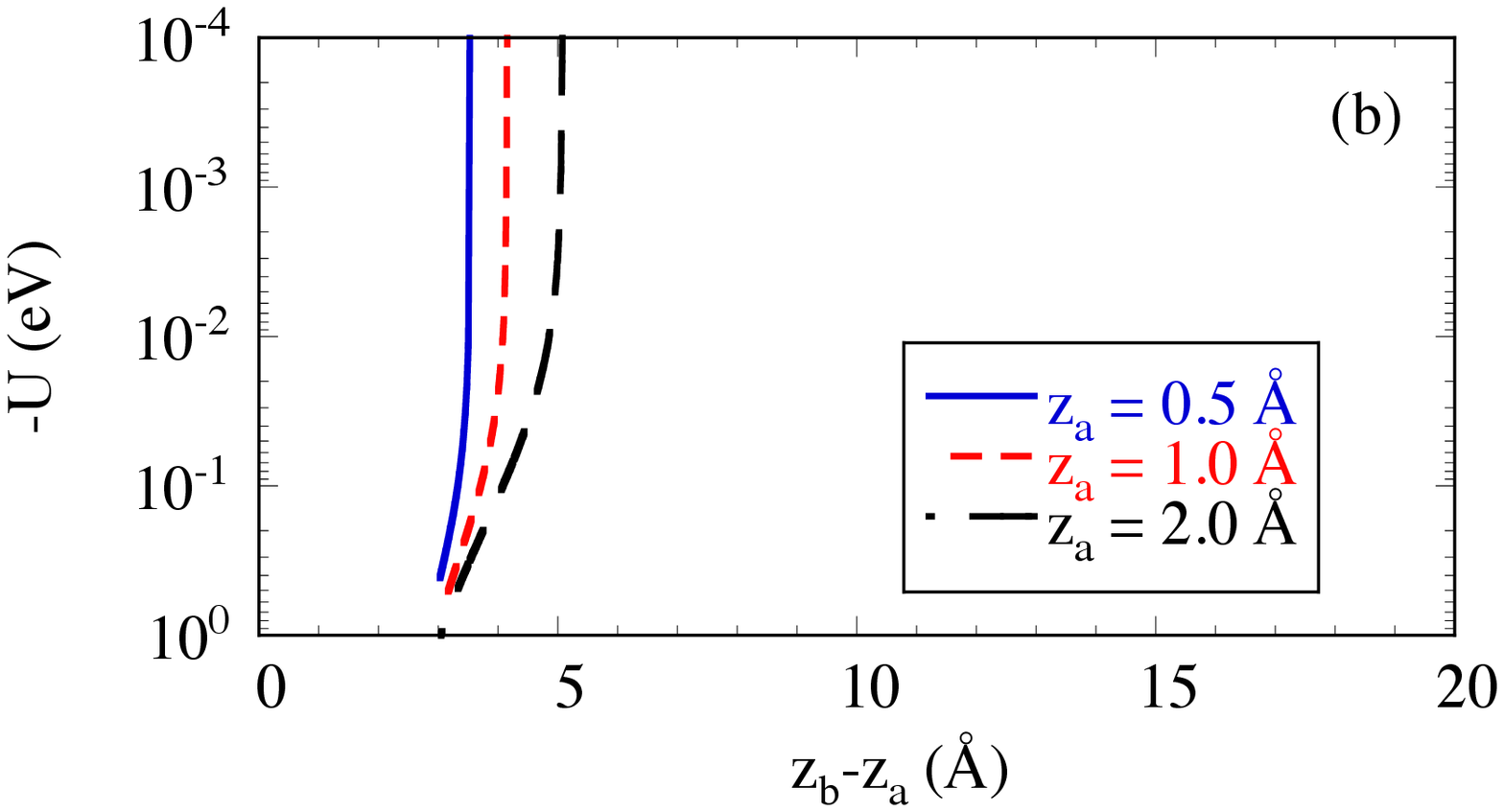}
\caption{(Color online)  The resonance interaction energy between two  isotropically excited  Li atoms situated one outside 
the other near a cellulose surface. The closest atom is at ${z_a}$ from the surface.   When the atoms are close to each other an attractive van der Waals interaction ($\propto \rho^{-6}$) dominates the resonance interaction. A van der Waals potential is also included in the DFT calculations. Atomistic DFT calculations  give the non-resonant bond enthalpy for this geometry to be 0.11 eV with a bond length of 2.81\,\AA.}
\label{figu4}
\end{figure}

\begin{figure}
\includegraphics[width=8cm]{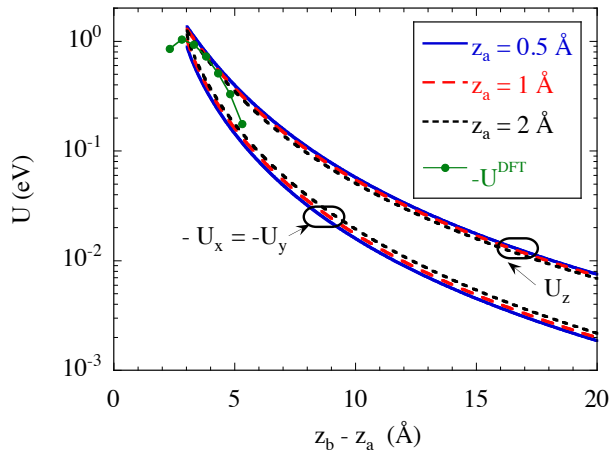}
\caption{(Color online) The three different contributions to the interaction potential in Figure\,\ref{figu4}. When the excitation is in the direction from one atom to the other a repulsion  (of the order 1 eV) follows that can break diatomic ${\rm{L}}{{\rm{i}}_2}$. For comparison we have added a curve (DFT curve) showing the attractive non-resonant interaction potential for two Li atoms near the surface as obtained from a DFT calculation, including a van der Waals functional.}
\label{figu5}
\end{figure}

In  Figs.\,\ref{figu4} and\,\ref{figu5} we consider the case with one Li atom adsorbed on a cellulose surface and the second Li 
atom situated directly outside the first atom.  Atomistic DFT calculations produce a non-resonant bond enthalpy of 0.11 eV for this 
geometry.  The non-resonant interaction energy, defined as the energy of the actual configuration relative to the energy when the outer atom is moved away to infinity, is larger (close to 1 eV). When the diatomic ${\rm{L}}{{\rm{i}}_2}$ is excited  in the direction defined by a line joining the 
two atoms there will be a strong  repulsive resonance interaction (above 1 eV) that could split the pair into two Li atoms.  Besides causing breaking of the bond the effect of the repulsive resonance interaction energy is to drive apart diatomic Li$_2$ for atom-atom distances beyond the bond length. Bond splitting can occur near surfaces and in nanostructures. More importantly the atomic Li is in the presence of a surface 
free to form either physical or chemical bonds with the cellulose surface and even diffuse into the cellulose layer structure. 
The calculated resonance interaction energy includes an attractive van der Waals potential ($\propto \rho^{-6}$). A similar term  is also included in the DFT evaluated non-resonant bond enthalpy which decreases fast with separation. It should be an important contribution to the non-resonant term for large separations.  For two atoms a distance of the order {4-5\,\AA} apart the resonance repulsion can overcome short range non-resonant interaction energies.

To conclude, we have  seen that resonance interaction between two Li atoms in an excited state near a cellulose surface gives rise to energies important enough to enlarge or even cause splitting of diatomic ${\rm{L}}{{\rm{i}}_2}$ molecules. 
For separations larger than the bond length atom pair breakage via repulsive resonance interaction can occur.  
In free space the bond enthalpy is as large as the 
upper limit of the resonance repulsion. There is therefore only a small probability to have bond breakage in free space.  Near a surface on the other hand bond breakage more easily occurs, and this breakage is enhanced when the atom pair is excited along the line connecting the two atoms. The repulsive resonance interaction is around 1 eV  for  ${\rm{L}}{{\rm{i}}_2}$ at close contact and decreases more slowly with separation compared to the attractive non-resonant bond enthalpy (which is around 0.1 eV at a cellulose surface).  Atomic Li can diffuse along the nanoporous cellulose and into the layered surface structure. 
It is possible that bond breaking via resonance repulsion can explain how atomic Li can be stored in cellulose nanostructures.

\begin{figure}
\includegraphics[width=8cm]{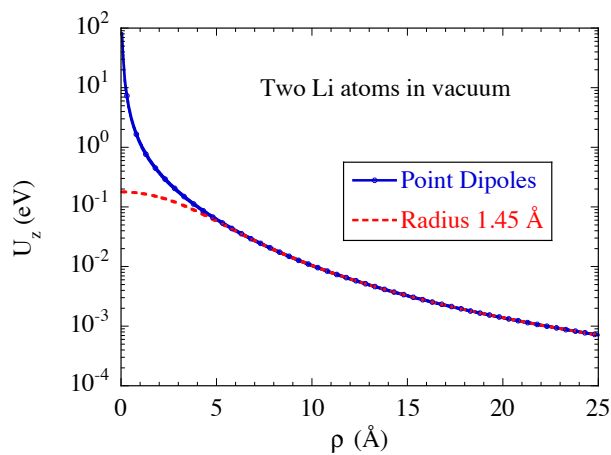}
\caption{(Color online) The resonance interaction  between two excited Li atoms in vacuum a distance $\rho$ apart (with the excitation in the direction between the two atoms).}
\label{figu6}
\end{figure}

Furthermore, in this work it is important to consider the fact that finite-atomic-size effects are negligible for separations larger than 
the bond length.\,\cite{Mah,Richardson}  This is seen in Fig.\,\ref{figu6} where we show the repulsive interaction for excited state diatomic ${\rm{L}}{{\rm{i}}_2}$  in free space  with the excitation in the direction joining the two atoms. The result for finite size atoms deviates from the point dipole approximation for separations less than the bond length. Such finite atomic size effects on the resonance interaction between atoms have not been considered 
in the past.

MB and CP aknowledge support from Swedish Research Council (Contract No. C0485101), STEM (Contract No. 34138-1), and The Research Council of Norway (Contract No. 221469/F20).  DH, WY, and CP acknowledge support from Swedish Research Council (Contract No. A0414101).



\begin{thebibliography}{10}

\bibitem{Sush} Sushkov A. O., Kim W. J., Dalvit D. A. R., and Lamoreaux S. K. , {\it Nature Physics}, {\bf 7} (2011) 230.
\bibitem{Zwol1} van Zwol P. J. and Palasantzas G. ,  {\it Phys. Rev. A} {\bf 81} (2010) 062502.
\bibitem{Feiler} Feiler A. A., Bergstr{\"o}m L., and Rutland M. W., {\it Langmuir} {\bf 24} (2008) 2274.
\bibitem{Hu} Hu L. , Wu H. , La Mantia F. , Yang Y. , Cui Y. , {\it ACS Nano} {\bf 4} (2010) 5843.
\bibitem{ChunJMC2012} Chun S.-J. , Choi E.-S. , Lee E.-H., Kim J. H.,  Lee S.-Y. and  Lee S.-Y., {\it J. Mat. Chem.} {\bf 22} (2012) 16618.
\bibitem{Nyst} Nystr{\"o}m G., Razaq A., Str{\o}mme M., Nyholm L. and  Mihranyan A., {\it Nano Lett.} {\bf 9} (2009) 3635.
\bibitem{Guo} Guo H., Mao R., Yang X.,  Wang S. and Chen J., {\it J. Power Sources} {\bf 219} (2012) 280.
\bibitem{Yue} Yue L.,  Zhong H. and Zhang L., {\it Electrochim. Acta} {\bf 76} (2012) 326.
\bibitem{Sherkunov} Sherkunov Y., {\it Phys. Rev. A} {\bf  75} (2007) 012705. 
\bibitem{Bostrom1} Bostr\"{o}m M.,  Longdell J. J.,  Mitchell D. J. and  Ninham B. W.,   {\it Eur. Phys. J. D} {\bf 22} (2003) 47. 
\bibitem{BostPRA2012}  Bostr\"{o}m M.,  Brevik I., Sernelius Bo E.,  Dou M.,  Persson C. and  Ninham B. W., {\it Phys. Rev. A} {\bf 86} (2012) 014701. 
\bibitem{Xie}  Xie L.,  Zhao L.,  Wan J.-l.,   Shao Z.-q.,  Wang F.-j. and  Lv S.-y., {\it J. Electrochem. Soc.} {\bf 159} (2012) A499.
\bibitem{Grimme} Grimme S., {\it J. Comp. Chem.} {\bf 27} (2006) 1787.
\bibitem{Nishy}  Nishiyama Y. ,  Langan P. and  Chanzy H., {\it J. Am. Chem. Soc.} {\bf 124} (2002) 9074.
\bibitem{Buck} Bucko T. ,  Tunega D.,  Angyan J. G. and Hafner J., {\it J. Phys. Chem. A} {\bf 115} (2011) 10097. 
\bibitem{Li} Li Y.,  Lin M. and   Davenport J. W., {\it J. Phys. Chem. C} {\bf 115} (2011) 11533.
\bibitem{Shish}  Shishkin M.,  Marsman M. and Kresse G., {\it Phys. Rev. Lett.} {\bf 99} (2007) 246403.
\bibitem{Gonz2010} Gonz\'{a}lez-Borrero P. P.,  Sato F.,  Medina A. N.,  Baesso M. L.,  Bento A. C.,
 Baldissera G.,   Persson C.,  Niklasson G.,  Granqvist C. G. and  Ferreira da Silva A., {\it Appl. Phys. Lett.} {\bf 96} (2010) 061909.
\bibitem{Gajd}  Gajdos M.,  Hummer K.,  Kresse G.,  Furthmuller J. and  Bechstedt F., {\it Phys. Rev. B} {\bf 73} (2006) 045112.
\bibitem{Expt1}  McQuarrie D. A. and  Simon J. D., {\it Physical Chemistry: A Molecular Approach} (University Science Books, Sausalito, 1997) 
\bibitem{Expt2}  Fournier R.,  Cheng J. B. Y., and  Wong A., {\it J. Chem. Phys.} {\bf 119} (2003) 9444.
\bibitem{Mah1974} Mahanty J. , {\it Il Nouvo Cimento B} {\bf 22} (1974) 110.
\bibitem{Mah} Mahanty J.  and  Ninham B. W., {\it Dispersion Forces} (Academic, London, 1976).
\bibitem{Pars}  Parsegian V. A., {\it Van der Waals forces: A handbook for biologists, chemists, engineers, and physicists}, (Cambridge University Press, New York, 2006). 
\bibitem{Ninhb}  Ninham B. W. and  Lo Nostro P., {\it Molecular Forces and Self Assembly in Colloid}, in Nano Sciences and Biology, (Cambridge University Press, Cambridge, 2010).
\bibitem{Buhmann1}  Safari H.,   Buhmann S. Y.,  Welsch D. -G. and  Trung Dung H.,  {\it Phys. Rev. A} {\bf 74} (2006) 042101.
\bibitem{Buhmann2} Buhmann S. Y. ,  Safari H.,  Trung Dung H. and  Welsch D. -G., {\it Optics and Spectroscopy} {\bf 103} (2007) 374.
\bibitem{McLone}  McLone R. R. and  Power E. A., {\it Mathematika} {\bf 11} (1964) 91.
\bibitem{WangPRA} Wang H., Li J., Wang X. T.,  Williams C. J.,  Gould P. L. and  Stwalley W. C., {\it Phys. Rev. A} {\bf 55} (1997) R1569.
\bibitem{Richardson}  Richardson D. D., {\it J. Phys. C: Solid State Phys.} {\bf 11} (1978) 3763.
\end{thebibliography}
\end{document}